\documentclass{jetpl}
\twocolumn

\usepackage{amsmath}
\usepackage{amssymb}
\usepackage{bm}
\usepackage{epsfig}
\usepackage{color}
\usepackage{graphicx}

\lat

\title{Multilayer graphene waveguides}
\rtitle{Multilayer graphene waveguides\ldots}

\sodtitle{Multilayer graphene waveguides}

\author{Daria\,A.\,Smirnova$^{a}$ \thanks{e-mail: daria.smirnova@anu.edu.au}, Ivan\,V.\,Iorsh$^{b}$, Ilya\,V.\,Shadrivov$^{a}$, and Yuri\,S.\,Kivshar$^{a,b} $}

\rauthor{D.\,A.\,Smirnova, I.\,V.\,Iorsh, I.\,V.,\,Shadrivov, and Yu.\,S.\,Kivshar}

\sodauthor{Smirnova, Shadrivov, Iorsh, Kivshar}

\address{ $^a$Nonlinear Physics Center, Research School of Physics and Engineering, Australian National University \\ Canberra ACT 0200, Australia \\~\\
$^b$National Research University of Information Technologies, Mechanics and Optics (ITMO), St. Petersburg 197101, Russia\\~\\}

\dates{\today}{*}

\abstract{We study dispersion properties of TM-polarized electromagnetic waves guided by a multilayer graphene metamaterial. We demonstrate that both dispersion and localization of the guided modes can be efficiently controlled by changing the number of layers in the structure. Remarkably, we find that in the long wavelength limit, the dispersion of the fundamental mode of the $N$-layer graphene structure coincides with the dispersion of a plasmon mode supported by a single graphene layer, but with $N$ times larger conductivity. We also compare our exact dispersion relations with the results provided by the effective media model.
}

\begin{document}
\def\dfrac{\displaystyle\frac}

\newcommand {\e}{{\rm e}}
\renewcommand {\d}{{\rm d}}
\renewcommand {\i}{{\rm i}}
\renewcommand {\Re}{\mathop{\mathrm{Re}}\nolimits}
\renewcommand {\Im}{\mathop{\mathrm{Im}}\nolimits}
\newcommand {\rot}{\mathop{\mathrm{rot}}\nolimits}
\newcommand {\divr}{\mathop{\mathrm{div}}\nolimits}
\newcommand {\grad}{\mathop{\mathrm{grad}}\nolimits}

\renewcommand {\phi}{\varphi}
\newcommand {\erf}{\mathop{\mathrm{erf}}\nolimits}

  \newcommand{\eL}{\varepsilon_\parallel}
  \newcommand{\eT}{\varepsilon_\perp}
  \newcommand{\eps}{\varepsilon}
\maketitle

Graphene plasmonics has attracted significant interest from the nanophotonics research community~\cite{Koppens_2011, Grigorenko_2012}, as it studies surface plasmon-polariton waves guided by an one-atom-thin graphene layer. These waves are characterized by the wavelength which is much smaller than the free space wavelength, and they can exist in the THz and far-infrared frequency ranges. Graphene plasmons were extensively studied theoretically~\cite{Koppens_2011,Newmode,theor2,theor3,theor4,Engheta,theor5}, and more recently plasmons in graphene were observed in experiment~\cite{Koppens_exp,Basovexp}.

One of the main obstacle impeding the efficient use of graphene in plasmonic devices is the difficulty of excitation of graphene surface plasmon modes, which is due to their deep subwavelength nature~\cite{RevLuo}. In this Letter, we suggest to employ multilayer graphene structures to overcome this difficulty. Multilayer graphene metamaterials have been studied previously~\cite{iorsh, Capolino}, and it has been shown that coupling of the surface plasmons at individual graphene sheets results in the emergence of the hyperbolic isofrequency contours, that can lead to a large density of electromagnetic states in these structures. Here we study the eigenmode dispersion of the multilayer graphene structures with a finite number of layers. We demonstrate that the field localization and modes' wavenumbers can be efficiently controlled by varying the number of layers in the structure, making such stacked graphene structures perspective for real optoelectronic and nanophotonic applications and the observation of strong nonlinear effects~\cite{Coupler_PRB, DissipSoliton_LPR}. Additionally, in contrast to the  metal-dielectric structures, the properties of multilayer graphene waveguides can be tunable by  means of the electrostatic doping of graphene or by applying an external magnetic field~\cite{iorsh_hybrid}.

\begin{figure}[!b]
\centerline{\includegraphics[width = 0.9\columnwidth]{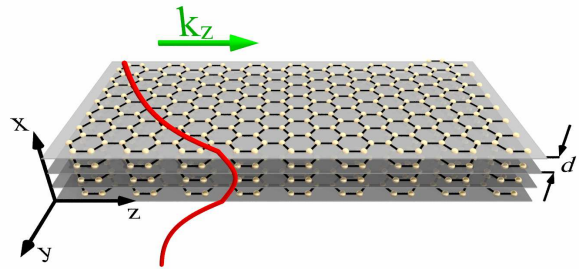}}
\caption{Figure 1. Geometry of the multilayer graphene waveguide. The broken curve shows schematically the profile of the continuous $z$-component of the electric field for the fundamental mode guided by the slab.}
\label{Fig_1}
\end{figure}

We consider a structure shown schematically in Fig.~\ref{Fig_1}. The graphene waveguide consists of a finite number of layers with deeply subwavelength spacing $d$ filled with dielectric material with permittivity $\eps$. The waveguide is sandwiched between two semi-infinite homogeneous dielectrics, the so-called substrate and superstrate, having dielectric constants of $\eps_1$ and $\eps_2$, respectively. Each of the $N$ graphene layers is placed at $x=md$, where $m \in [0, N-1]$. The capping medium and substrate medium therefore occupy the half-spaces $x<0$ and $x>(N-1)d$, respectively.
We assume harmonic time-dependence for modes propagating along $z$ axis $\exp(-i \omega t)$ so that the propagation along $z$-axis is described by the multiplier  $\exp( i k_z z)$, where $k_z=\beta k_0$ is the propagation constant, $k_0=\omega/c$ is the wavenumber in a free space, and $\beta$ is the normalized wavenumber.
Next, we derive the dispersion relation by employing the matrix method~\cite{Yeh, Born}. In the regions $md\leq x \leq (m+1)d$, the transverse profile of the magnetic field can be presented in the form, 
\begin{equation}
H_y(x) = H^{m}_{+}e^{-\kappa x} + H^{m}_{-}e^{\kappa x}\:,
\end{equation}
where $\kappa = (k_z^2 - k_0^2\eps)^{1/2}$ is the transverse wavenumber.
From the Maxwell equations, we find the $z$-component of the electric field continuous at the graphene layers, given by $E_z(x)=({i \kappa}/{k_0 \eps}) H_y(x)$.
Since we are looking for localized guided modes vanishing for large $|x|$, we present the field outside the waveguide in the following form
\begin{equation}
H_y(x<0)\! =\! H^0_{-}e^{\kappa_1 x}\! , \quad H_y\left(x>(N-1)d\right)\! =\! H^{N\!G}_{+}e^{-\kappa_2 x}\!,
\end{equation}
where $\kappa_{1,2} = (k_z^2 - k_0^2\eps_{1,2})^{1/2}$.
To study waves in a multilayer structure, it is convenient to use the transfer matrix method which links the field amplitudes in the adjacent periods, $H^{m+1}_{\pm}$ and $H^{m}_{\pm}$
\begin{equation}
\begin{pmatrix} H^{m+1}_{+} \\ H^{m+1}_{-} \end{pmatrix} = \hat T \times \left( \begin{array}{c} H^{m}_{+} \\ H^{m}_{-}  \end{array} \right)\:,
\end{equation}
where the transfer matrix $\hat T=\hat P \hat G$ is a product of the matrices describing the boundary conditions at the graphene layer
\begin{equation}
\hat G=
\left( {\begin{array} {cc}
{ 1 -  {2 \pi i\sigma(\omega) \kappa}/ {\omega \eps}} &
{   {2 \pi i\sigma(\omega) \kappa}/ {\omega \eps}}  \\
{ - {2 \pi i \sigma(\omega) \kappa}/ {\omega \eps}}   &
{ 1 +  {2 \pi i \sigma(\omega) \kappa}/ {\omega \eps}}  \\
 \end{array} } \right)\:,
\end{equation}
where $\sigma(\omega)$ is the frequency-dependent surface conductivity of a single layer of graphene, and $\hat P$ being the propagation matrix of a dielectric layer defined as
\begin{equation}
\hat P=
\left( {\begin{array} {cc}
e^{-\kappa d} &
0 \\
0 &
e^{\kappa d} \\
 \end{array} } \right)\:.
\end{equation}

In our calculations, we neglect losses and assume $\hbar \omega < 1.67 \mu \: (\Im \sigma  > 0)$, where $\mu$ is the chemical potential, also taking conductivity for highly doped or gated graphene ($k_B T \ll \mu$) in the form~\cite{RevLuo}
\begin{equation}
\sigma(\omega) = \displaystyle{\frac{ie^2}{\pi \hbar}\left[\frac{\mu }{\hbar \omega } + \frac{1}{4} \text{ln}\frac{(2\mu -\hbar \omega)}{(2\mu + \hbar \omega)}\right]}\:, \\ \label{eq:sigma}
\end{equation}
where $e$ is the charge of electron, $k_B$ is the Boltzmann constant, and $T$ is the temperature.

For an infinite periodic structure, we employ the Bloch theorem, $H^{m+1}_{\pm}=H^{m}_{\pm}e^{iK_{B}d}$, and obtain the dispersion of Bloch waves in the form
\begin{equation}
\cos({K_{B}}d) = \cosh(\kappa d) - \displaystyle{\frac{\kappa}{2\eps}\frac{4 \pi \sigma(\omega)}{i c k_0 }}\sinh(\kappa d)\:.\label{Bloch1}
\end{equation}
If $d$ is large enough (or, equivalently, $k_z \rightarrow \infty$), the above expression approaches the dispersion relation for surface p-polarized plasmons supported by a single graphene layer surrounded by dielectrics with permittivity $\eps$~\cite{JablanPRB}:
\begin{equation}
\frac{2\eps} {\kappa} = \frac{4 \pi \sigma(\omega)}{i c k_0}\:. \label{eq:DispEq1L}
\end{equation}

The matrices of the boundary conditions for the outermost graphene layers can be written as
\begin{align}
\hat{G}_1=\frac{1}{2}\begin{pmatrix} 1+\mathcal{Z}_1 & 1-\mathcal{Z}_1  \\ 1-\mathcal{Z}_1^{\text{*}} & 1 + \mathcal{Z}_1^{\text{*}} \end{pmatrix},
\end{align}
\begin{align}
\hat{G}_{N\!G}=\frac{1}{2}\begin{pmatrix} 1+\mathcal{Z}_{N\!G} & 1-\mathcal{Z}_{N\!G}  \\ 1-\mathcal{Z}_{N\!G}^{\text{*}} & 1 + \mathcal{Z}_{N\!G}^{\text{*}} \end{pmatrix},
\end{align}
where
\begin{align}
\mathcal{Z}_1=\frac{\kappa_1\eps}{\kappa \eps_1}-\frac{4\pi}{c}\frac{i\sigma(\omega)\kappa_1}{k_0\eps_1},\\
\mathcal{Z}_{N\!G}=\frac{\kappa\eps_2}{\kappa_2 \eps}-\frac{4\pi}{c}\frac{i\sigma(\omega)\kappa}{k_0\eps}.
\end{align}

A linear relationship between the field amplitudes on both sides of the multilayer structure can be written in the matrix form 
\begin{equation}
\begin{pmatrix} H^{N\!G}_{+} \\ 0 \end{pmatrix} = \hat M \times \left( \begin{array}{c} 0 \\ H^{0}_{-}
\end{array} \right)\:,
\end{equation}
where the matrix $\hat M$ is obtained sequentially multiplying the matrices $\hat G$ and $\hat P$: $\hat M = \hat G_{N\!G}  (\hat P\hat G)^{N-2} \hat P\hat G_1$. By setting $m_{22}=0$, we obtain the dispersion relation for localized modes, from which we can find the wavenumber for a given frequency numerically. Once we found the wavenumber, we then calculate the corresponding wavefunction using the matrix relation for the amplitudes in the adjacent periods. The physical origin of the modes is similar to that in other coupled systems, and $N$ interacting graphene sheets will support $N$ non-degenerate plasmon modes originating from coupling of plasmons of individual graphene layers.
\begin{figure}[!b]
\centerline{\includegraphics[width = 0.9\columnwidth]{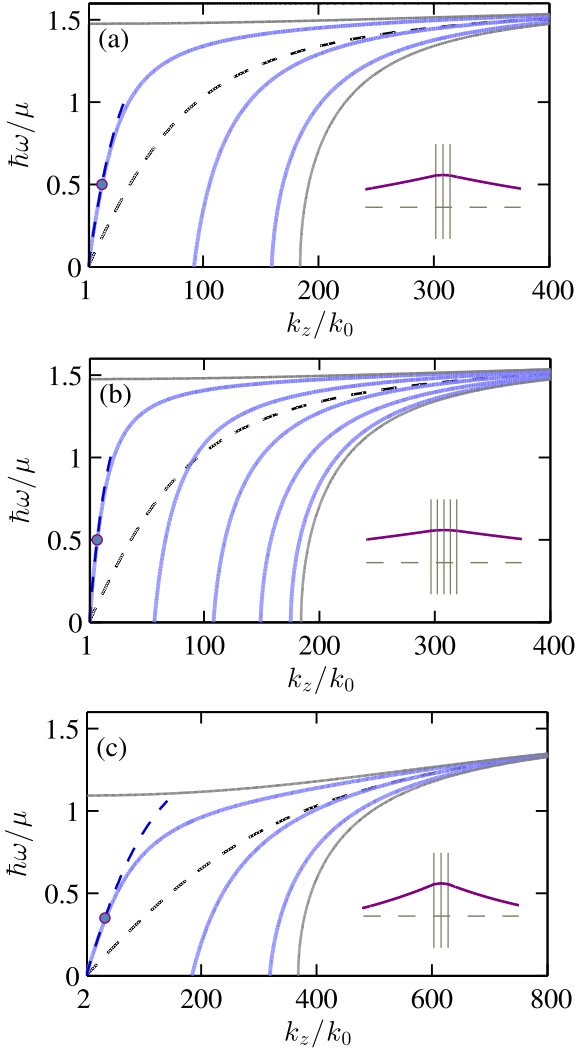}} 
\caption{Figure 2. Dispersion of the eigenmodes of the multilayer graphene waveguides consisting of 3 (a,c) and 5 (b) layers. The surrounding media and interlayer spacers are made from the same material with dielectric permittivity $\eps = 1$ (a,b) and $\eps = 4$ (c). Period of the structure is $d=8$ nm in all the cases. Gray solid lines show the boundaries of the allowed band for the localized propagating solutions inside the graphene slab. Black dashed line shows the dispersion of the surface plasmon for a single graphene layer; blue dashed line corresponds to the dispersion of a surface plasmon localized at the two-dimensional conducting layer with permittivity $N\sigma$. Insets show the electric field profile $|E_z|$ for the fundamental mode at frequency $\hbar\omega=0.5\mu$ (a,b) and $\hbar\omega=0.35\mu$ (c).}
\label{fig2}
\end{figure}
\begin{figure}[!b]
\centerline{\includegraphics[width = 0.9\columnwidth]{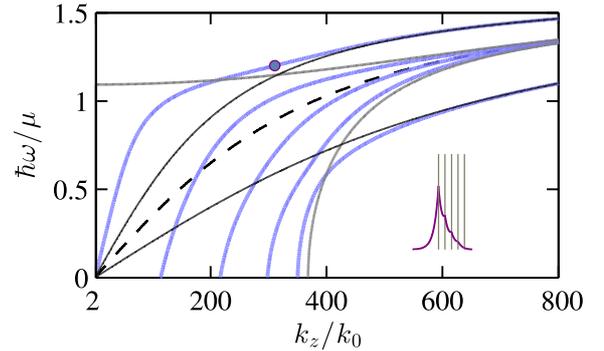}} 
\caption{Figure 3. Dispersion of the asymmetric multilayer graphene waveguide, $\eps_1=1$, $\eps_2=10$, $\eps=4$, $d=8$ nm. Gray solid lines show the boundaries of the allowed band for the localized propagating solutions inside the graphene slab. The dispersion branches that cross the boundary of the allowed band correspond to the surface Tamm states. Inset shows the profile of the electric field $|E_z|$ for the Tamm state for the parameters depicted with the point at the dispersion branch at $\hbar \omega = 1.2 \mu$.}
\label{fig3}
\end{figure}

All the possible cases of the dielectric permittivities distributions can be divided by symmetry into symmetric ($\eps_1=\eps_2$) and asymmetric ($\eps_1 \neq \eps_2$) cases. In the case $\eps_1=\eps_2=\eps$, all the modes lie in the allowed band, whereas in all other cases branches can span in the forbidden band where the Bloch number has an imaginary part and, thus, the respective modes are strongly localized in the vicinity of the waveguide boundary.

In the most general case when all three dielectric constants ($\eps$, $\eps_1$ and $\eps_2$) are different, for large separations (or, equivalently, $k_z \rightarrow \infty$) there exist $(N-2)$ degenerate states with an asymptote given by Eq.~(\ref{eq:DispEq1L}) and 2 non-degenerate states with asymptotes
\begin{equation}
\frac{\eps} {\kappa} + \frac{\eps_{1,2}} {\kappa_{1,2}} = \frac{4 \pi \sigma(\omega)}{i \omega}\:. \label{eq:DispEq2L}
\end{equation}
Being mostly confined near the layers $x=0$ and $x=(N-1)d$, two latter modes are related to the  surface Bloch waves localized near the edges of the multilayer waveguide, and they can be attributed to the so-called Tamm states.

Next, we consider a symmetric case $\eps_1=\eps_2=\eps$. For $N=2$ and $\eps=\eps_1=\eps_2$, we have $\hat M = \hat G \hat P \hat G $, and immediately obtain the dispersion relation for the symmetric and antisymmetric modes guided by a graphene double-layer structure, studied earlier in Refs~\cite{Coupler_PRB, buslaev_JETP}:
\begin{equation}\label{disp_two}
1 + \dfrac {2 \pi i \sigma(\omega) \kappa} {\omega \eps} \left( 1 \pm e^{-\kappa d} \right) =0\:.
\end{equation}
Importantly, for small spacing between the layers ($\kappa d \ll 1$), the \textit{symmetric} mode dispersion [positive sign in Eq.~(\ref{disp_two})] coincides with the dispersion of a single plasmon (\ref{eq:DispEq1L}), where conductivity is double that of a single graphene layer. It means that for the fundamental symmetric mode, adding two graphene layers effectively doubles the conductivity of graphene.

For $N\geq 3$, transfer matrix is $\hat M = \hat G (\hat P \hat G)^{N-1} $. Using the Tchebychev identity, the dispersion relation for the $N$-layer waveguide can be analytically simplified, namely,
\begin{equation}
t_{21} U_{N-2}(a) g_{12 }+ \left[ t_{22} U_{N-2}(a) - U_{N-3} (a)\right] g_{22}=0\:, \label{eq:Chebyshev}
\end{equation}
where $t_{ij}$, $g_{ij}$ are the elements of the matrices $\hat T$ and $\hat G$, $U_k(a)$ are the Tchebychev polynomials with the argument $a = (t_{11}+t_{22})/2$. We look for the case of closely spaced layers, and linearize Eq.~(\ref{eq:Chebyshev}) with respect to the small parameter $\kappa d$ to obtain
\begin{align}
1+\frac{2\pi i\kappa N\sigma(\omega)}{\omega\eps}+(N-1)\kappa d =0,
\end{align}
where the third term can be omitted due to smallness of $\kappa d$. This leads us to the equation for the surface plasmon dispersion~(\ref{eq:DispEq1L}), where the single graphene conductivity is replace by $N$ times larger conductivity, $N\sigma(\omega)$. Figure~\ref{fig2} shows the dispersion of the eigenmodes of the multilayer graphene waveguide for different parameters. We can see that for the fundamental mode in the low frequency region, where $\kappa d \ll 1$, the dispersion of the mode is well described by the dispersion of the plasmon localized at the graphene layer with permittivity $N\sigma$ (these dispersions are shown with blue dashed lines). The wavenumbers of such plasmons are significantly smaller, than those of the single layer, and therefore they should have longer propagation distances, being easier to excite.

If we now consider the asymmetric waveguide,  $\eps_1\neq\eps_2$, we observe the emergence of the surface electromagnetic states, with the dispersion lying outside the allowed band region as shown in figure~\ref{fig3}. As was mentioned above, this leads to a finite imaginary part of the Bloch wavevector and localization of the mode at one of the interfaces of the structure, as shown in the inset of Fig.~\ref{fig3}.
\begin{figure}[!t]
\centerline{\includegraphics[width = 0.9\columnwidth] {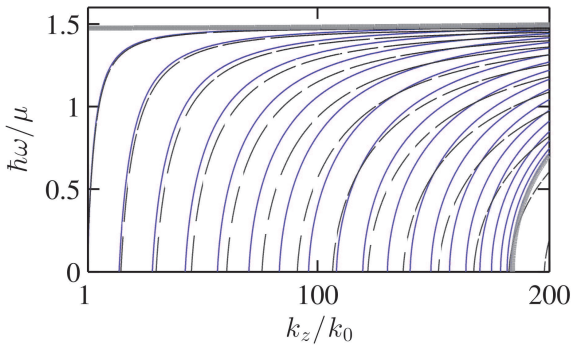}} 
\caption{Figure 4. Comparison between the dispersion curves of the eigenmodes obtained from the exact solution (solid lines) and by means of the effective model (dashed lines), for $N=20$, $d=8$ nm, $\eps_1=\eps_2=\eps=1$.}
\label{fig4}
\end{figure}
\begin{figure}[!b]
\centerline{\includegraphics[width = 0.9\columnwidth]{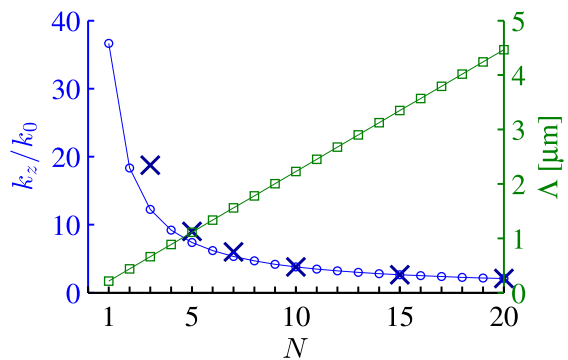}} 
\caption{Figure 5. Dependence of the normalized wavenumber of the fundamental guided mode $k_z/k_0$ (circles) and the effective width of the mode $\Lambda$ (squares) on the number of layers in the structure, solid curves are guides for eye. The parameters are $d=8$ nm, $\eps_1=\eps_2=\eps=1$, $\hbar\omega/\mu=0.5$, $\mu=0.1$ eV. Crosses show the result of the effective medium model. }
\label{fig5}
\end{figure}

Now we compare the obtained eigenmode dispersion with the results provided by the effective medium model. Within this model, the multilayer structure is described as an uniform hyperbolic medium with dielectric permittivity tensor components defined as~\cite{iorsh}: $\eps_{yy}=\eps_{zz}=\eps+4i\pi\sigma(\omega)/(ck_0d)$, $\eps_{xx}=\eps$.
General dispersion equation for the hyperbolic waveguide can be written in the form:
\begin{align}
&\cos(k_x D)(k_x \kappa_1 \eps_2 \eps_{zz}+ k_x \eps_{zz} \eps_1\kappa_2)+ \nonumber \\
&\sin(k_x D) (\eps_{zz}^2\kappa_1\kappa_2-k_x^2\eps_1\eps_2)=0,
\end{align}
where $k_x =[\eps_{zz}k_0^2-k_z^2(\eps_{zz}/\eps_{xx})]^{1/2}$ and $D=(N-1)d$. Figure~\ref{fig4} shows the comparison between the effective model and the exact solution obtained by the transfer matrix method. We notice a good agreement between the two approaches in the limit $k_xd \ll \pi$.

We would like to emphasize that, by varying the number of layers in the multilayer graphene waveguide, we can effectively control the wavenumber of the fundamental guided mode. When the wavenumber of the mode grows, it becomes harder to excite the mode optically from the vacuum. Moreover, another important property of the waveguide is the effective mode width $\Lambda$  which is a sum of the actual waveguide thickness and double the localization length of the waveguide mode outside the waveguide, $\Lambda = (N-1)d+2/\kappa$. 

Figure~\ref{fig5} shows the dependence of the values $k_z/k_0$ and $\Lambda$ on the number of layers. As was mentioned above, at the chosen frequency  the normalized wavenumber of the wave decreases as $1/N$, so that we can easier excite the mode by increasing the number of layers. Accordingly, the effective waveguide thickness grows linearly with the number of layers. We can also observe, that as we increase the number of layers, the mode dispersion is better described within the effective media model (shown with the crosses).

In conclusion, we have studied the dispersion properties of the plasmonic modes guided by multilayer graphene structures. We have revealed that the localization of the fundamental mode can be substantially controlled by varying the number of layers in the graphene structure, which can serve as an additional parameter for optimizing designs in graphene-based nanophotonics. We have demonstrated that by using multilayer graphene structures one can control efficiently 
the degree of localization of plasmon modes, as well as their group and phase velocities.

{\em Acknowledgments.} This work was supported by the Government of the Russian Federation (grant 074-U01) and the Australian National University.

\end{document}